# High-resolution spectroscopy of proximity superconductivity in finite-size quantized surface states

Lucas Schneider[1†*], Christian von Bredow[1‡], Howon Kim[1‡], Khai That Ton[1‡], Torben Hänke[1], Jens Wiebe[1] and Roland Wiesendanger[1]

[1]Department of Physics, University of Hamburg, D-20355 Hamburg, Germany.
† Present address: Department of Physics, University of California at Berkeley, Berkeley, CA, USA
‡ These authors contributed equally

*E-mail: lucas.schneider@physnet.uni-hamburg.de

**Abstract**
**Adding superconducting (SC) electron pairing via the proximity effect to pristinely non-superconducting materials can lead to a variety of interesting physical phenomena. Particular interest has recently focused on inducing SC into two-dimensional surface states (SSs), potentially also combined with non-trivial topology. We study the mechanism of proximity-induced SC into the Shockley-type SSs of the noble metals Ag(111) and Cu(111) grown on the elemental SC Nb(110) using scanning tunneling spectroscopy. The tunneling spectra exhibit an intriguing multitude of sharp states at low energies. Their appearance can be explained by Andreev bound states (ABS) formed by the weakly proximitized SSs subject to lateral finite-size confinement. We study systematically how the proximity gap in the bulk states of both Ag(111) and Cu(111) persists up to island thicknesses of several times the bulk coherence length of Nb. We find that even for thick islands, the SSs acquire a gap, with the gap size for Cu being consistently larger than for Ag. Based on this, we argue that the SC in the SS is not provided through direct overlap of the SS wavefunction with the SC host but can be understood to be mediated by step edges inducing electronic coupling to the bulk. Our work provides important input for the microscopic understanding of induced superconductivity in heterostructures and its spectral manifestation. Moreover, it lays the foundation for more complex SC heterostructures based on noble metals.**

**Main**
Combining materials with different individual properties into heterostructures has proven to be a viable route to enable new functionalities[1]. For instance, non-SC materials can inherit SC pairing from a nearby superconductor via the process of Andreev reflection at the interface[2–4]. This effect is known as proximity induced superconductivity and leads to a gapped energy region in the material's spectrum. Local probes such as scanning tunneling microscopy (STM) turned out to be invaluable to study lateral proximity effects, e.g. into monolayer Pb[5–7] or graphene[8] because of their ability to correlate high-resolution tunneling spectroscopy with atomically resolved information. Recently, the realm of topological materials has led to an increased interest in the proximity effect, as superconductivity in topological bands[9–12] or magnetic materials[13–16] is expected to be unconventional. Many proposals rely on induced SC pairing into surface states (SSs) that are well decoupled from the bulk, thus requiring mono- or few-layer growth of a material on superconducting surfaces, which can be challenging experimentally. Here, we present a high-resolution STM study of the interplay between lateral confinement and proximity superconductivity in the SSs of Ag(111) and Cu(111). These Shockley-type SSs have been shown to act like a non-interacting two-dimensional electron gas (2DEG), which can be confined into artificial[17,18] or naturally occurring[19] nanostructures. As we show below, we find evidence for a SC gap opening in the SSs even for islands with thicknesses on the order of tens of nanometers. We argue that this gap is opened by an indirect proximity effect via surface-bulk scattering by the most frequent defects which are step edges in the metal.

We deposited Ag and Cu respectively onto oxygen reconstructed Nb(110) while keeping the crystal at a temperature of 550K. Both Ag[17,20,21] and Cu form islands in the fcc(111) orientation [see Fig. S1 in the Supplementary Material[22]] following a Stranski-Krastanov growth mode [Figs. 1(a) and 1(b)], resulting in islands of varying height and area on top of a wetting layer (WL). In this work, we use SC Nb tips for an enhanced energy resolution[17,23] at junction temperatures between 320mK and 4.5K. Approximately, the presence of the SC tip gap $\Delta_t \approx$ 1.35meV shifts all spectroscopic features to higher energies by an amount $\pm\Delta_t$ [orange marked zone in Figs. 1(c) and 1(d)]. Tunneling spectroscopy measured on the respective WL of Ag and Cu at mK temperatures reveals clear signs of Superconductor-Insulator-Superconductor (SIS) tunneling [Figs. 1(c) and 1(d)]. Prominent tunneling occurs between the tip's coherence peak and the first de-Gennes-Saint-James (dGSJ) resonance[3] of the sample (marked by red arrows). These peaks have recently been identified in STS experiments on proximitized Au films[24,25] and behave closely analogous to the coherence peak of the superconducting layer in thin NS bilayers[25,26]. Therefore, we identify their energy with the value of the respective bulk gap $\Delta_b$ in the normal metal layer[25,27]. Its measured values are close to the Nb bulk gap of $\Delta_0 = 1.5$ meV for the wetting layer, indicating a decent NS junction transparency and Fermi surface matching. We traced the estimated energetic position of these outermost dGSJ resonances for Ag(111) and Cu(111) islands of varying thicknesses. The results are shown in Fig. 1(g) and indicate the gradual suppression of the SC bulk gap induced in the N layer with the layer thickness. A large value of $\Delta_b$ persists even in islands significantly thicker than the bulk coherence length of the underlying Nb superconductor ($\xi_{Nb}$ = 38 nm[28]), evidencing an enhanced renormalized coherence length in a clean normal metal given by $\xi_N = \hbar v_F/(2\pi k_B T)$, where $v_F$ is the Fermi velocity in the N layer[29] (note that mismatched Fermi velocity between the SC and N layers complicates this expression[30]). We model the decay of $\Delta_b$ with island thickness by calculating the position of the first dGSJ resonance[3,24–26] in the normal metal for a different effective coherence lengths $\xi_N$ by solving the equation $2d/\xi_N \cos[\phi] = (n\pi + \phi)$ for $n = 1$ and $\Delta_b = \Delta_0 \cos[\phi]$ [26]. Four examples are shown as the dashed lines in Fig. 1(g). Although, as expected, this simple model is not able to describe the data perfectly well, it demonstrates that $\xi_N$ is clearly larger than $\xi_{Nb}$. Notably, the decay is similar for both Ag and Cu, which can be understood as a consequence of the similar Fermi velocities $v_F \approx 1 \cdot 10^6$ m/s in both noble metals[31,32] yielding $\xi_N \approx 270$ nm at $T$ = 4.5 K. Nevertheless, the gap values found for Cu tend to be slightly smaller than for Ag. We speculate that this is a result of the slightly more disordered growth of the Cu wetting layer, which might result in a reduced junction transparency for Cu/Nb(110) compared to Ag/Nb(110). The long effective coherence length in the noble metal layers verifies that our samples are in the clean limit. It opens avenues for superconducting heterostructures with a full gap even for thick islands.

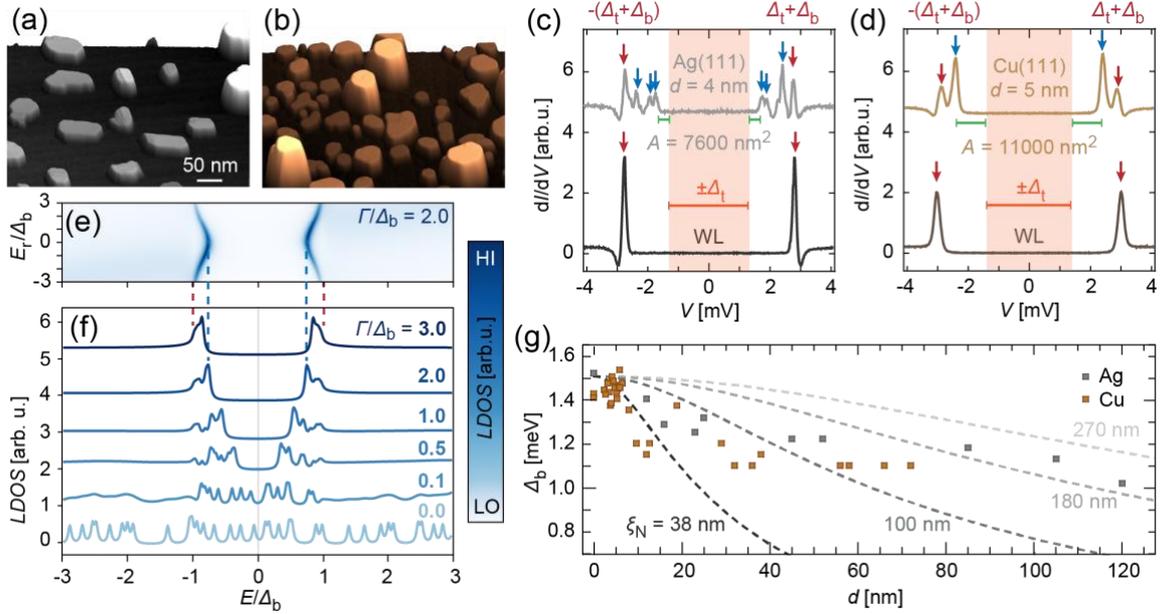

Fig. 1. (a), 3D rendering of a constant-current STM topography showing multiple Ag islands of up to 11 nm thickness grown on Nb(110). (b), 3D rendering of a constant-current STM topography showing multiple Cu islands of up to 18 nm thickness grown on Nb(110). (c), d$I$/d$V$ spectroscopy measured on the pseudomorphically grown Ag wetting layer (WL) and on a 4 nm thick Ag(111) island. The red arrows mark the estimated position of the bulk coherence peaks, while the blue arrows mark sub-gap states. (d), d$I$/d$V$ spectroscopy measured on the Cu wetting layer (WL) and on a 5 nm thick Cu(111) island. In panels c and d the spectra are vertically offset for clarity, the tip gap region $\pm\Delta_t$ is indicated in orange, and the green bars mark the region around zero energy without sub-gap peaks. (e), Calculated LDOS($E$) for a single resonance level at energy $E_r$. The coupling $\Gamma$ to the SC is set to be $2\Delta_b$. (f), Calculated LDOS($E$) for a randomly shaped island of area $A = 5000$ nm$^2$ with $\Delta_0 = 1.35$ meV. The bulk gap and surface state gaps for $\Gamma/\Delta_b = 3.0$ are marked by red and blue lines, respectively. An artificial broadening has been added as an imaginary part $i \cdot \delta E = 0.03\Delta_b$ to the energy $E$. (g), Position of the bulk coherence peaks vs. island thickness for both Ag and Cu. The dashed lines show the expected shift of the first dGSJ resonance[3] for different values of the effective coherence length in the noble metal $\xi_N$, see text. Here, we set $\Delta_0$ = 1.5 meV, corresponding to the bulk gap of Nb(110). Parameters: $V$ = 1000 mV, $I$ = 20 pA, $T$ = 4.2K (panel a), $V$ = 1000 mV, $I$ = 10 pA, $T$ = 4.5K (panel b), $V_{stab}$ = 5 mV, $I$ = 1 nA, $V_{mod}$ = 0.05 mV, $T$ = 0.32K (panel c), $V_{stab}$ = 6 mV, $I$ = 0.3 nA, $V_{mod}$ = 0.03 mV, $T$ = 0.35K (panel d).

Besides the dGSJ states, we find a plethora of additional resonances at sub-gap energies for the fcc(111) islands [blue arrows in Figs. 1(c) and 1(d)]. While for Cu(111), these states gather close to $\Delta_b$ and leave a comparably large region without sub-gap conductance (marked by green bars), they are found at almost all sub-gap energies in Ag(111). However, even for Ag, there is a small region without peaks around $E_F$ (located at a bias of $V = \pm\Delta_t$, marked by green bars). Concerning the energies, the sub-gap spectra look remarkably particle-hole symmetric around $E_F$, i.e. there is a sub-gap peak at energy $-E$ for every peak at $+E$. However, there is a particle-hole asymmetry for the intensities of these peaks. In the following, we demonstrate that these peaks can be understood as a series of the recently discovered Machida-Shibata-States (MSSs)[17,33], spin-degenerate Andreev bound states induced by laterally confined SS modes in contact to the bulk SC substrate. Furthermore, we will show that the regions without peaks marked with green bars can be associated with a SC gap $\Delta_{SS} < \Delta_b$ in the SS modes.

For a single non-interacting resonance mode, the energy dependent local density of states (LDOS) has been derived in Ref. [17] to be

$$\text{LDOS}(E) = -\frac{1}{\pi}\text{Im}\left[\frac{E + E_r + \frac{\Gamma E}{\sqrt{\Delta_b^2 - E^2}}}{E^2\left(1 + \frac{2\Gamma}{\sqrt{\Delta_b^2 - E^2}}\right) - E_r^2 - \Gamma^2}\right] \quad (1)$$

with the resonance level's energy $E_r$, the bulk gap $\Delta_b$ and a coupling between surface- and bulk states $\Gamma$. The LDOS for a single level for fixed $\Gamma$ and varied $E_r$ is shown in Fig. 1(e). For $|E_r| \gg \Delta_b$, the MSSs merge with the dGSJ resonance at energies $E \approx \pm\Delta_b$. For smaller $|E_r|$, their energies move into the gap, but never reach zero as the proximity effect into the resonance level opens up a gap in the spectrum[17].

In contrast to Ref. [17], we study islands with large area $A$ in this work. Accordingly, not only one but multiple confined eigenstates are expected simultaneously near the Fermi energy $E_F$. In a confined 2DEG, we expect the average level spacing between confined states $E_r$ to be $\delta E_r(A) = \frac{2\pi\hbar^2}{A \cdot m^*}$ with the effective mass of the SS electrons of $m^* = 0.4 m_e$ for Ag(111) and Cu(111)[19,34,35]. We simulated spectra of randomly shaped islands with a given $A$ and increasing $\Gamma$ by summing up the LDOS for all respective individual resonance levels given by Eq. (1). One example for an island of $A = 5000$ nm$^2$ is shown in Fig. 1(f). For zero coupling to the bulk $\Gamma/\Delta_b \simeq 0$ (bottom most spectrum), eigenstates are randomly distributed in energy leading to arbitrary finite-size gaps having an average size of $\delta E_r/2$. The observed spectroscopic gap of the islands now crucially depends both on the induced SC pairing and on this finite-size quantization of the SS modes. However, the finite-size gaps are not necessarily symmetric with respect to the Fermi level $E_F$ because the metallic surface state does not have particle-hole symmetry. For finite coupling (spectra with increasing $\Gamma/\Delta_b$), the resonances outside the superconducting bulk gap become smeared out because of scattering into the metallic bulk states[36,37]. In turn, the resonances inside the gap remain very sharp due to the lack of bulk states to scatter into. The proximity effect acting onto each of the resonance levels[17] opens a small superconducting gap around $E_F$ and makes the spectra more particle-hole symmetric regarding the energies of the resonances. This scenario is similar to the observed sub-gap state distribution on Ag(111) where we find sub-gap states everywhere except within a small region around $E_F$ and the low-energy states are particle-hole symmetric [Fig. 1(c)]. Correspondingly, we identify this inner region as the induced gap in the surface state modes. For strong coupling (topmost two spectra), the sub-gap states induced by a random distribution of $E_r$ bunch at the extrema of the curve in Fig. 1(e), i.e. at $\pm\Delta_b$ (marked by red dashed lines) and at the proximity gap for a given $\Gamma$ (blue dashed lines), giving the impression of only two pairs of peaks for $\Gamma/\Delta_b \geq 2.0$. This is remarkably reminiscent of the data for Cu(111) in Fig. 1(d), indicating that the sub-gap peaks in fact consist of multiple MSSs that are not all individually resolved. In this limit, the system can be interpreted as a two-band superconductor with two different gap sizes[20]. Our work thus bridges the gap between the physics of the proximity effect acting on single, strongly confined states[17] and the physics of two-band superconductors[20,23].

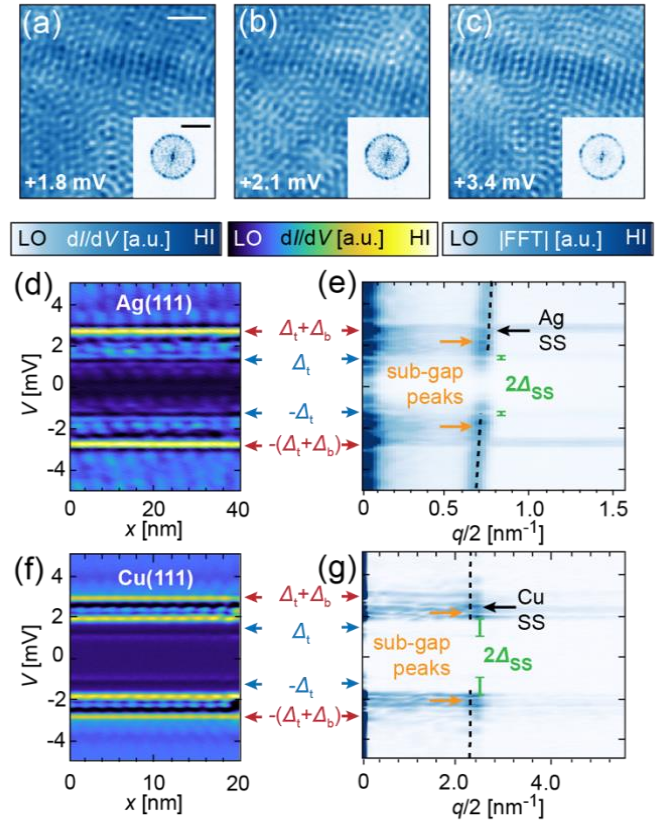

Fig. 2. (a)-(c), d$I$/d$V$ maps measured on Ag(111) at sub-gap (a,b) and out-of-gap (c) energies showing interference of the SS electrons. Insets: FFT of the real space maps showing a circle of isotropic scattering near $q = q_F$. The scale bar in the FFT corresponds to $q/2 = 1$ nm$^{-1}$. (d), Line of d$I$/d$V$ spectra measured across the same island. The values of the tip and sample gaps $\Delta_b$ and $\Delta_t$ are indicated by arrows on the right side. Note that these measurements were conducted at elevated temperatures therefore, thermally activated tunneling is occurring even at bias voltages between $\pm\Delta_t$[38]. (e), Energy dependent radial average of the FFTs shown in the inset of panel a. The black dashed line represents the known dispersion for the Ag(111) SS with $m^* = 0.40 m_e$ and $E_0 = -50$ meV[17,39]. (f), Line of d$I$/d$V$ spectra measured across a terrace on Cu(111). The values of the tip and sample gaps $\Delta_b$ and $\Delta_t$ are indicated again. (g), Energy dependent FFT of panel f. The black dashed line represents the expected dispersion for the Cu(111) SS with $m^* = 0.48 m_e$ and $E_0 = -400$ meV[17,39]. The green bars in panels e and g mark the region in the spectra around zero energy without sub-gap states, which we identify with $\Delta_{SS}$. Parameters: $V_{stab}$ = 10 mV, $I$ = 1 nA, $V_{mod}$ = 0.1 mV, $T$ = 4.2K (constant height mode, panels a-e), $V_{stab}$ = 5 mV, $I$ = 1 nA, $V_{mod}$ = 0.03 mV, $T$ = 1.7K (panels f,g).

In order to prove their surmised relation with the Ag and Cu SSs, we map the spatial dependence of the sub-gap states. In Figs. 2(a)-2(c), we present d$I$/d$V$ maps at sub-gap and out-of-gap energies measured on Ag(111). All three examples exhibit very similar quasiparticle interference (QPI) patterns of the SSs, as evidenced by the Fourier transform of the maps [insets in Fig. 2(a)] showing the circular pattern known for QPI on fcc(111) noble metals[39,40]. In a complementary measurement, a line of d$I$/d$V$ spectra measured across a Ag(111) island [Fig. 2(d)] reveals that the sub-gap states are following the same periodicity as the d$I$/d$V$ features outside the superconducting gap. This can be seen in the energy dependent Fourier transform of spectroscopic maps [Fig. 2(e)] showing the dispersion of the Ag(111) SS, and closely matching previous results[39,40] for energies outside $\pm\Delta_t$. Between $\pm\Delta_t$ and $\pm(\Delta_t + \Delta_b)$, i.e. inside the sample's bulk gap, it can be seen that the sub-gap peaks show prominent intensity at $q = q_F$ [orange arrows in Fig. 2(e)]. This demonstrates how the MSSs are tied to the Ag SSs. The scenario can be thought of like in a two-gap superconductor where density of states from the band with a smaller gap can exist inside the larger gap. However, it has never been shown that these features in the density of a two-band superconductor are spectroscopically this sharp and protected by the gapped bulk. Very similar features are observed for Cu(111), i.e. spatial oscillations of the sub-gap states with $q = q_F$ of the SS [Figs. 2(f) and 2(g), c.f. Fig. S1 in the Supplementary Material[22]]. The most important difference is a larger gap $\Delta_{SS}$ of the peaks around $E_F$ (marked by green bars). For both metals, we observe a suppression of the LDOS in the FFT at $q = q_F$ within an interval $\pm\Delta_{SS}$ around $\Delta_t$ (however, this is barely visible for Ag, we refer to Fig. 3 for a clearer representation). This indicates that the suppression is indeed a gap $\Delta_{SS}$ in the SS band.

The origin of the induced gap $\Delta_{SS}$ cannot be a direct proximity effect from the underlying Nb(110) substrate because the decay length of the SSs in the direction perpendicular to the surface is only a few nm[20]. Therefore, we argue that the superconductivity has to be induced indirectly by surface-bulk scattering into the proximitized noble metal bulk states[17,41]. Surface-bulk scattering in noble metals is known to be enhanced by impurities and step edges[36,41,42], and to be stronger for Cu(111) than for Ag(111)[36,43]. We can investigate this effect by studying islands with different areas and consequently a different effective step edge density. Fig. 3(b) shows a line of d$I$/d$V$ spectra along the topography in Fig. 3(a) across terraces of the same

Ag(111) island. While the difference in thickness between the terraces is negligible with respect to the effective coherence length $\xi_N$ [see Fig. 1(g)], they cover a considerably different area $A$. A variety of sub-gap states can be seen, however, there always appears a gap around $E_F$ where no peaks can be found. The magnitude of this gap clearly changes when the tip moves to different terraces but is constant within a single terrace. We have measured the gap magnitude $\Delta_b$ for tens of different islands and terraces for both Ag(111) and Cu(111) [Fig. 3(c)]. Notably, there is a trend of a decreasing gap size with increasing island area $A$ for both metals, while the data points for Ag fall consistently below those for Cu, consistent with the weaker surface-bulk coupling for Ag. When interpreting the spectroscopic gaps, we have to consider the interplay between finite-size quantization and induced SC mentioned above. Therefore, we plot the average finite-size gap magnitude $\delta E_r(A)/2$ together with the data for the measured spectral gaps in Fig. 3(c) (gray dashed line). As the measured data points lie systematically above the expectation value for finite-size gaps, we conclude that the gaps are of induced SC type and that they represent the SC gap in the surface states $\Delta_{SS}$. The fact that the gap size decreases with $A$ indicates that the ratio between terrace interior and edge is indeed determining the surface-bulk coupling, and correspondingly the indirect proximity effect[17]. A related result is found for a sample with a higher impurity density induced by sputtering the Ag film (see Fig. S2 in the Supplementary Material[22]) where $\Delta_{SS}$ is strongly enhanced, in agreement with the idea that impurity scattering facilitates the proximity effect. Thus, imperfect growth of multilayers on superconductors can boost the induced gaps by reduced surface state lifetime[41,44]. For realistic applications of heterostructures maintaining quantum coherence, there will be a delicate trade-off between coupling to the SC and SS lifetime.

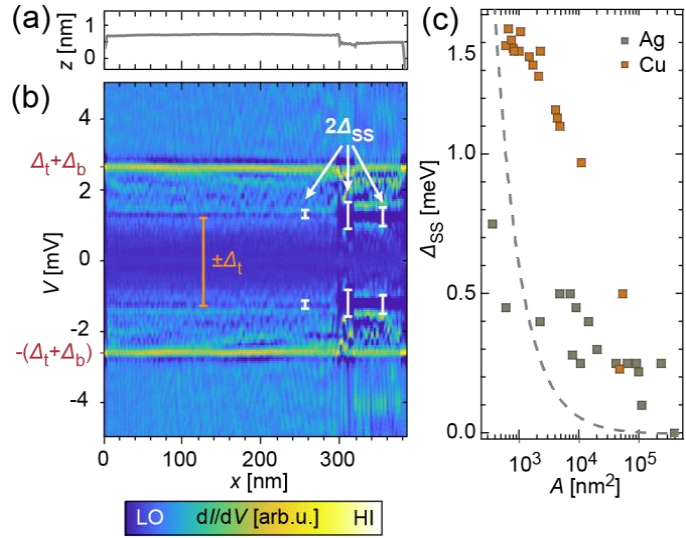

Fig. 3. (a), Topographic profile across three atomic terraces on a Ag(111) island. (b), d$I$/d$V$ line profile across the same three terraces. The energies of the tip gap $\Delta_t$, the sum of tip and Ag bulk state gap ($\Delta_t$ + $\Delta_b$) as well as the surface state gap $\Delta_{SS}$ are marked respectively. (c), Dependence of the surface state gap $\Delta_{SS}$ on the terrace area $A$. The gray dashed line represents the expected average spacing $\delta E_r(A)/2$

of the lowest eigenenergies in a 2DEG without superconducting pairing and thus the finite-size gaps. Parameters: $V_{stab}$ = 5 mV, $I$ = 1 nA, $V_{mod}$ = 0.05 mV, $T$ = 4.5K

We believe that our microscopic understanding of the proximity effect in nanoscale islands helps interpreting other complex proximity systems, e.g. such investigated in Ref. [45] where gapless surface states coexist with gapped bulk bands. In another direction, surface-bulk coupling leading to linewidth broadening was a strongly limiting factor in previous studies of confined surface states[46]. Our proximitized structures circumvent this problem by gapping out the bulk of the normal metal, leading to resonance levels with widths down to the resolution limit of our measurement (approx. 50µeV). This could enable high-resolution studies of artificial lattices or quantum chaos in billiards with high quantum numbers, which has turned out to be impossible in conventional noble-metal platforms because of level broadening[37]. Most importantly, our work opens up possibilities to study well characterized heterostructures like thin films on noble metals[47,48], magnetic adatoms[49] and molecules or graphene nanoribbons[50,51] and their interplay with superconductivity in an atomically precise platform.

**Data availability**
The authors declare that the data supporting the findings of this study are available within the paper and its supplementary information files.

**Code availability**
The analysis codes that support the findings of the study are available from the corresponding authors on reasonable request.

**Acknowledgements**
L.S., J.W., and R.W. gratefully acknowledge funding by the Cluster of Excellence 'Advanced Imaging of Matter' (EXC 2056 - project ID 390715994) of the Deutsche Forschungsgemeinschaft (DFG). R.W., H.K. and T.H. acknowledge funding of the European Union via the ERC Advanced Grant ADMIRE (project No. 786020). We thank Roberto Lo Conte, Thore Posske and Ioannis Ioannidis for helpful discussions.

**Author contributions**
L.S. and J.W. conceived the experiments. L.S., C.v.B., K.T.T., H.K. and T.H. performed the measurements and analyzed the experimental data. L.S. and C.v.B. prepared the figures, L.S. wrote the manuscript. All authors contributed to the discussions and to correcting the manuscript.

**Competing interests**
The authors declare no competing interests.